**Comment on "Long Range Charge Ordering in Magnetite below the Verwey Transition".**

(Submitted to Phys. Rev. Lett.)

In a recent letter, J. P. Wright et al [1] have investigated the crystal structure of magnetite below the Verwey transition temperature ($T_V$) by means of X–ray and neutron powder diffraction. The presence of two classes of $FeO_6$ octahedra in magnetite (expanded, B1 and B4 and compressed, B2 and B3) is the main experimental result of the crystal refinement. The authors conclude a valence of 2.4 and 2.6 for expanded and compressed octahedra, respectively, by applying the empirical bond valence sums method (BVS) and using the average Fe–O bond length at each site as physical parameter. The observation of these two different octahedra is the origin of their claiming for long–range charge ordering (CO) in magnetite below $T_V$.

In this comment, we will show that the interpretation given by J. P. Wright et *al.* is unsupported by their own crystallographic data. Instead, these data demonstrate the lack of atomic long–range CO in magnetite below the Verwey transition [2–4].

J. P. Wright et *al.* have shown that magnetite crystallizes in a Cc supercell below $T_V$ in agreement with a previous single crystal study [2]. The full refinement of such supercell is not available so far and the reported structural analysis only concerns a subcell (with symmetry constraints) of the true crystal cell but the authors have proposed two possible CO models (class I or II). The authors state that their analysis is equivalent to averaging the true crystal structure but this average fails to show the presence of $Fe^{3+}$ and $Fe^{2+}$ in magnetite. In fact, the average Fe–O distance for B1+B4 and B2+B3 octahedra are 2.070 Å and 2.046 Å, respectively, very different from the expected values for either class I CO, B1+B4($4Fe^{2+}$) = 2.16 Å and B2+B3($4Fe^{3+}$) = 2.025 Å, or class II CO model, B1+B4($3Fe^{2+}+Fe^{3+}$) = 2.126 Å and B2+B3 ($3Fe^{3+}+Fe^{2+}$) = 2.059 Å. Therefore, experimental data do not support the existence of $Fe^{3+}$ or $Fe^{2+}$ octahedra. On the other hand, if we consider that the refined Fe–O distances [1] are very close to the real distances in the true cell (as expected), the authors' interpretation is in clear contradiction with the BVS model. None of the averaged experimental <d(Fe–O)> distances of B1,B2, B3 and B4 octahedron[1] agrees with those expected for $Fe^{2+}$ or $Fe^{3+}$ octahedral ions[6]. The authors try to overcome this strong discrepancy with the BVS model suggesting that this model does not work in CO systems (see ref. 21, 25 or 26 in ref. 1). However, none of these papers show experimental evidence of atomic CO, and most important, the BVS model would lose its validity. The valence renormalization criterion, reported in table I, is not coherent as well.

Furthermore, the proposed fractional charge disproportionation (20%) between compressed and expanded $FeO_6$ octahedra should only be considered as a maximum limit. The reasons for this affirmation are:

1) The single crystal refinement given by Iizumi et *al.* [2] did not show two classes of $FeO_6$ octahedra in different refinements. For instance, the average Fe–O distances for the $FeO_6$ octahedra are (Pmca model): $B_1$= 2.071 Å, $B_2$= 2.039 Å, $B_3$= 2.053 Å and $B_4$= 2.061 Å.

2) Even though powder diffraction provides data more reliable than single crystal diffraction (anyhow, 2160 reflections were measured in the single crystal study [2]), the precision in the charge determination should also take into account the BVS sensitivity. The average absolute deviation in the linear relationship between bond–valence parameters is around 0.017 Å for oxides [5]. Accordingly, the total sensitivity (including experimental errors) should be of the order of 0.022 Å. This value is comparable to the Fe–O differences found between B1–B4 and B2–B3 octahedra.

3) The refined distances are the average of the true crystal structure so it is difficult to differentiate two kinds of octahedra. The standard deviation of Fe–O distances for an individual octahedron is larger than the difference of the average distance among different octahedra. For instance, B1 and B4 have a <d(Fe−O)> = 2.07 Å with a standard deviation of Fe−O distances σ= 0.022 Å whereas B1 and B3 have <d(Fe−O)> = 2.0466 Å and σ= 0.038 Å (Note that the difference between both <d(Fe−O)> is 0.0234 Å).

In conclusion, we can state that the reported crystallographic data either by powder [1] or by single crystal diffraction [2] are not compatible with the existence of $Fe^{3+}$ and $Fe^{2+}$ octahedra. It is noteworthy that the standard deviation for the whole octahedral distances distribution [1] (<d(Fe−O)> = 2.058 Å and σ= 0.033 Å) is much smaller than the difference in distance between reference $Fe^{3+}$ and $Fe^{2+}$ octahedra (0.135 Å)[6]. Therefore, we can conclude that the octahedral iron in magnetite shows an intermediate valence state below the Verwey transition as it has been previously concluded from x−ray resonant scattering experiments[4].

Joaquin Garcia*, Gloria Subías[#], Javier Blasco* and M. Grazia Proietti*.

* ICMA, CSIC− Universidad de Zaragoza. 50009 Zaragoza, Spain

# ESRF, BP220, 38043 Grenoble, France